\documentclass[iop,apj]{emulateapj}

\shorttitle{The lesser role of starbursts for star formation  at $\lowercase{z}=2$}
\shortauthors{Rodighiero et al.}
\slugcomment{Accepted in Astrophysical Journal Letter}

\begin{document}

\title{The lesser role of starburstsfor star formation at $\lowercase{z}=2$}
\thanks{Herschel is an ESA space observatory with science instruments provided by European-led Principal Investigator consortia and with important participation from NASA.}

\author{
G. Rodighiero\altaffilmark{1,}\altaffilmark{2},
E. Daddi,\altaffilmark{3},
I. Baronchelli\altaffilmark{1},
A. Cimatti\altaffilmark{4},
A. Renzini\altaffilmark{5},
H. Aussel\altaffilmark{3},
P. Popesso\altaffilmark{6},
D. Lutz\altaffilmark{6},
P. Andreani\altaffilmark{7},
S. Berta\altaffilmark{6},
A. Cava\altaffilmark{8},
D. Elbaz\altaffilmark{3},
A. Feltre\altaffilmark{1},
A. Fontana\altaffilmark{9},
N. M. F{\"o}rster Schreiber\altaffilmark{6},
A. Franceschini\altaffilmark{1},
R. Genzel\altaffilmark{6},
A. Grazian\altaffilmark{9},
C. Gruppioni\altaffilmark{10},
O. Ilbert\altaffilmark{11},
E. Le Floch\altaffilmark{3},
G. Magdis,\altaffilmark{3},\altaffilmark{12},
M. Magliocchetti\altaffilmark{13},
B. Magnelli\altaffilmark{6},
R. Maiolino\altaffilmark{9},
H. McCracken\altaffilmark{14},
R. Nordon\altaffilmark{6},
A. Poglitsch\altaffilmark{6},
P. Santini\altaffilmark{9},
F. Pozzi\altaffilmark{4},
L. Riguccini\altaffilmark{3},
L.J. Tacconi\altaffilmark{6},
S. Wuyts\altaffilmark{6},
G. Zamorani\altaffilmark{10}
}

\altaffiltext{1}{Dipartimento di Astronomia, Universita' di Padova, Vicolo dell'Osservatorio 3, I-35122,  Italy}
\altaffiltext{2}{email: giulia.rodighiero@unipd.it}
\altaffiltext{3}{Laboratoire AIM, CEA/DSM-CNRS-Universit{\'e} Paris Diderot, IRFU/Service d'Astrophysique, B\^at.709, CEA-Saclay, 91191 Gif-sur-Yvette Cedex, France.}
\altaffiltext{4}{Dipartimento di Astronomia, Universit\`a di Bologna, via Ranzani 1, I-40127 Bologna, Italy}
\altaffiltext{5}{INAF-Osservatorio Astronomico di Padova, Vicolo dell'Osservatorio 2, I-35122 Padova, Italy}
\altaffiltext{6}{Max-Planck-Institut f\"{u}r  extraterrestrische Physik, Postfach 1312, 85741 Garching, Germany}
\altaffiltext{7}{ European Southern Observatory, Karl-Schwarzschild-Str. 2, 85748 Garching, Germany}
\altaffiltext{8}{Departamento de Astrof\'{\i}sica, Facultad de CC. F\'{\i}sicas, Universidad Complutense de Madrid, E-28040 Madrid, Spain} 
\altaffiltext{9}{INAF-Osservatorio Astronomico di Roma, via di Frascati 33, 00040 Monte Porzio Catone, Italy}
\altaffiltext{10}{INAF-Osservatorio Astronomico di Bologna, Via Ranzani 1, I-40127, Bologna, Italy}
\altaffiltext{11}{Laboratoire d'Astrophysique de Marseille, Universit«e de Provence, CNRS, BP 8, Traverse du Siphon, 13376 Marseille Cedex 12, France}
\altaffiltext{12}{Department of Physics, University of Oxford, Keble Road, Oxford OX1 3RH}
\altaffiltext{13}{INAF-IFSI, Via Fosso del Cavaliere 100, I-00133 Roma, Italy}
\altaffiltext{14}{Institut d'Astrophysique de Paris, UMR7095 CNRS, Universit\' e Pierre et Marie Curie, 98 bis Boulevard Arago, 75014 Paris, France}

\begin{abstract}

Two main modes of star formation are know to control the growth of galaxies: a relatively steady one 
in disk-like galaxies, defining a tight star formation rate (SFR)-stellar mass sequence, and a starburst
 mode in outliers to such a sequence which is generally interpreted as driven by merging.  Such starburst 
galaxies are rare but have much higher SFRs, and it is of interest to establish the relative importance 
of these two modes.  PACS/Herschel observations over the whole COSMOS and GOODS-South fields, in 
conjunction with previous optical/near-IR data, have allowed us to accurately quantify for the first time 
the relative contribution of the two modes to the global SFR density in the redshift interval $1.5<z<2.5$,
 i.e., at the cosmic peak of the star formation activity.  The logarithmic distributions of galaxy SFRs at fixed stellar 
mass are well described by Gaussians, with starburst galaxies representing only a relatively minor deviation 
that becomes apparent for SFRs more than 4 times higher than on the main sequence.  Such starburst galaxies 
represent only 2\% of mass-selected star forming galaxies and account for only 10\% of the cosmic SFR density 
at $z\sim2$. 
Only when limited to SFR$>1000M_{\odot}/yr$, off-sequence sources significantly contribute to the SFR density ($46\pm20$\%).
We conclude that merger-driven starbursts play a relatively minor role for the formation of 
stars in galaxies, whereas they may represent a critical phase towards the quenching of star formation and 
morphological transformation in galaxies.  
                                   
\end{abstract}


\section{Introduction}
Star forming galaxies follow a tight correlation between their stellar
mass ($M_*$) and star formation rate (SFR), defining a {\it main
sequence} (MS) that has been recognized in the local Universe
(Brinchmann et al. 2004; Salim et al. 2007; Peng et al. 2010), as well
as at intermediate redshifts $0.5<z<3$ (Noeske et al.  2007; Elbaz et
al. 2007; Daddi et al. 2007; Pannella et al. 2009; 
Rodighiero et al. 2010a, Karim et al. 2011), and beyond (Daddi et
al. 2009; Gonzalez et al. 2010). 
With SFR $\propto M_*^{\alpha}$, the slope $\alpha$ can
differ substantially depending on sample selection and the 
   procedures for measuring SFR and $M_*$, with values in the above
literature ranging from $\sim 0.6$ to $\sim 1$. Moreover, its
normalization rapidly rises from $z=0$ to $z\sim2$--2.5 as $(1+z)^{\sim3.5}$, then
flattening all the way to the highest redshifts (Daddi et al. 2007,
Rodighiero et al. 2010a; Karim et al. 2011).
Slope and normalization of the SFR$-M_*$ relation play a crucial role in the
growth of galaxies and in the evolution of their mass function
(Renzini 2009; Peng et al. 2010, 2011).
Observations of the CO
molecular gas content of MS galaxies indicate that their star
formation efficiency does not depend strongly on cosmic epoch to
$z\sim2$, with the SFR increase being due to higher molecular gas
fractions (Daddi et al. 2008; 2010a; Tacconi et al. 2010; Geach et
al. 2011).

On the other hand, outliers are known to exist, with very high
specific SFR (SSFR) compared to normal MS galaxies, such as local
ultra-luminous infrared galaxies (ULIRG, Sanders \& Mirabel 1996, Elbaz et al. 2007), 
and at least some of the submm-selected galaxies (SMG)
at $1<z<4$ (Tacconi et al. 2008; Daddi et al. 2007; 2009; Takagi et
al.  2008).  What does differentiate such active behemoths from the
dominant MS population?  Outliers are galaxies in which the 
SSFR has been boosted by some event, possibly a
major merger (e.g., Mihos \& Hernquist 1996; di Matteo et al. 2008;
Martig et al. 2010; 
Bournaud et al. 2011), as suggested by local
ULIRGs and SMGs being dominated by gas-rich major mergers (e.g. Sanders et
al. 1988). Indeed, MS galaxies and outliers appear
to be in different star formation regimes: a quasi--steady,
long--lasting mode for disks and a more rapid, starburst mode in major
mergers or in the densest SF regions (Daddi et al. 2010b; Genzel et
al. 2010). 

So far it has been unclear by which of these two modes most of the
stars in galaxies were formed. While MS galaxies are optically thin in
the UV (Daddi et al. 2005; 2007), MS outliers are
generally optically thick (Goldader et al. 2002, Chapman et al. 2005) 
and far-IR observations are required to reliably
derive their SFRs. The PACS camera (Poglitsch et al. 2010) onboard
Herschel (Pilbratt et al. 2010) now allows for the first time to obtain deep far-IR
observation probing SFRs down to MS levels for typical galaxies with
$M_*\sim10^{10}M_\odot$ at $z\sim2$, i.e., at the peak epoch of the
cosmic SFR density and of the space density of SMGs (Chapman et
al. 2005).  In this letter we combine wide area PACS observations of
the COSMOS field with deeper data in the GOODS field, both taken as a
part of the PEP survey (Lutz et al. 2011), and obtain a first accurate
estimate of the relative role of MS and outlier galaxies on the
formation of stars at $z\sim2$.  
Red and dead (passive) galaxies, though, exist at these cosmic epochs, and form 
a separate sequence below the MS of star-forming galaxies.
Their contribution will be ignored in this Letter.
We use a Salpeter IMF and a standard WMAP cosmology.

\begin{figure*}[!ht]
\includegraphics{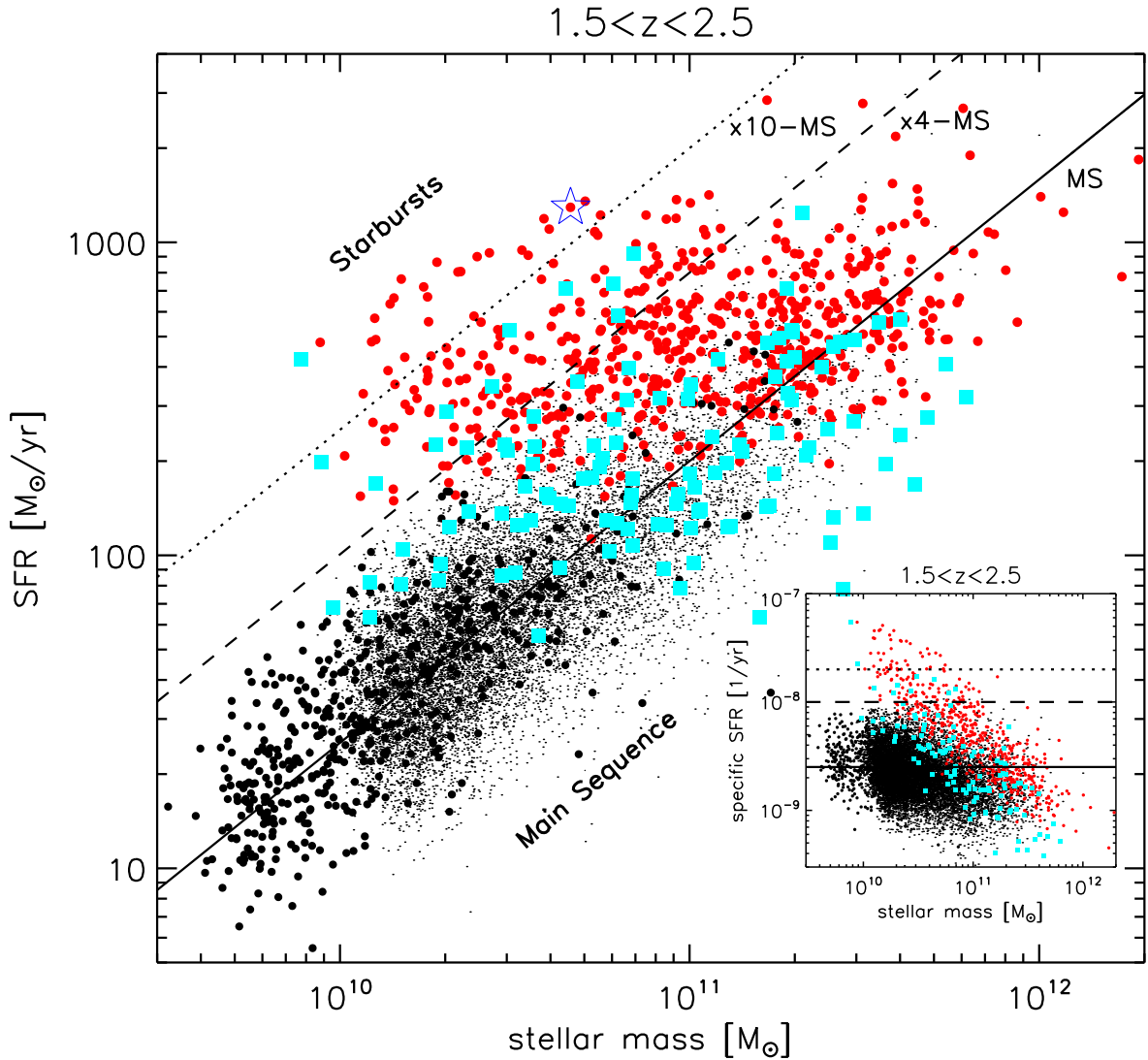}
\caption{Stellar mass -- Star Formation Rate relation at $1.5<z<2.5$. 
We use four main samples:  the "shallow" PACS-COSMOS sources 
(red filled circles), the deeper PACS-GOODS South (cyan squares), the
BzK-GOODS sample (black filled circles) and the BzK-COSMOS sources
(black dots).  The solid black line indicates the main sequence (MS) for
star-forming galaxies at $z\sim2$ defined by Daddi et al. (2007),
while the dotted and dashed lines mark the loci 10 and 4 times above
the MS (along the SFR axis), respectively.
The star
indicates the PACS source detected by Aztec at 1.1mm in the COSMOS field.
In the smaller inset, we show the same information as in the main
panel, however here the stellar mass is presented as a function of the
SSFR. 
}
\label{selection}
\end{figure*}

\begin{figure*}[!ht]
\centering
\includegraphics[width=16cm]{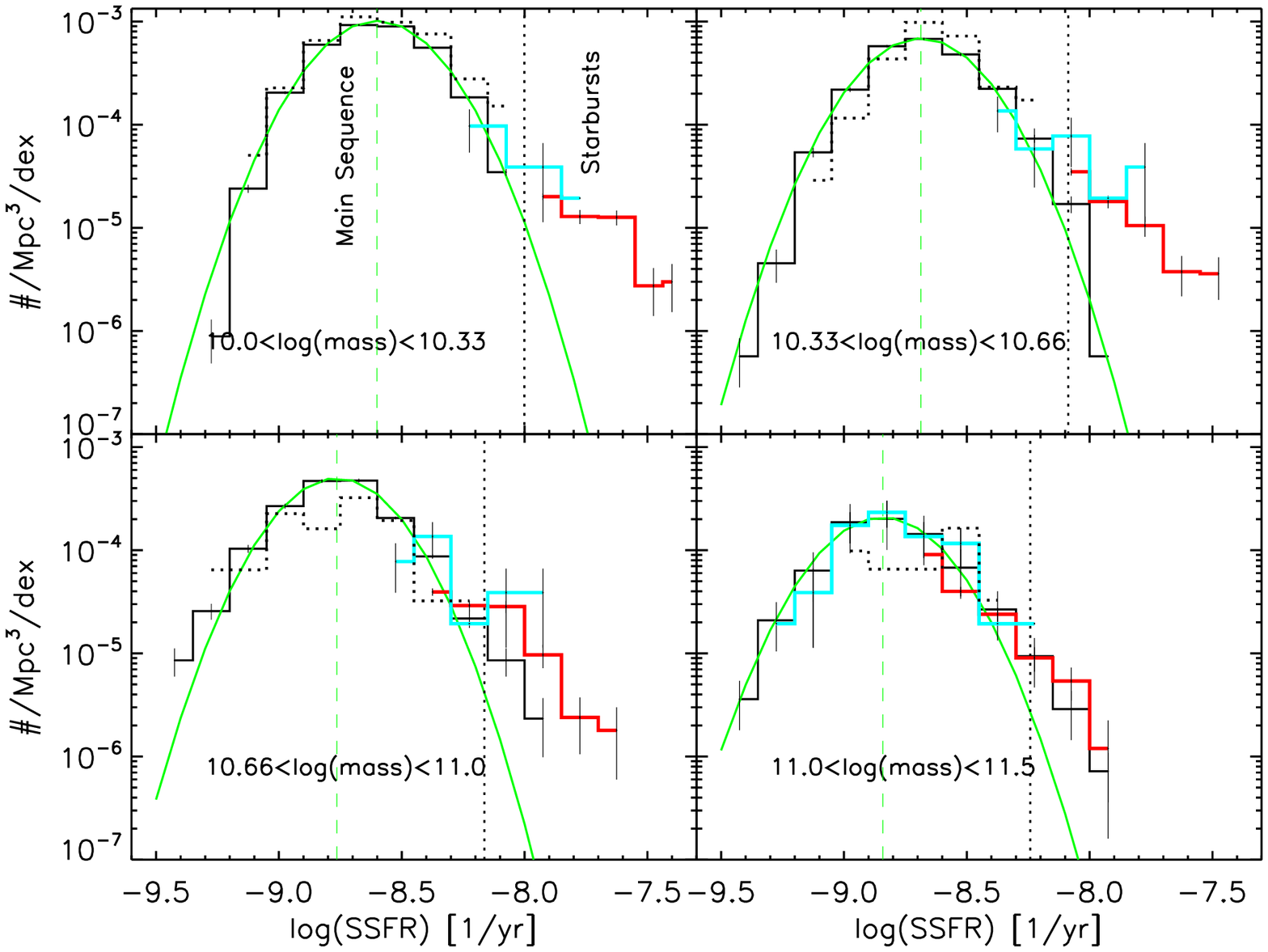}
\caption{Distribution of the SSFR for the four samples, corrected by
the corresponding comoving volumes and after accounting for volume and
selection incompleteness, splitted in four mass bins. The
dotted-black, solid-black, cyan and red histograms correspond to the
BzK-GOODS, BzK-COSMOS, PACS-GOODS South and PACS-COSMOS
data-sets. Error bars are Poisson. The
green curves show the Gaussian fits to the black-solid histograms.
The green dashed vertical lines mark the vertex of each best-fit
Gaussian, and the dotted black vertical lines +0.6 dex above show 
our choice to separate ON- and
OFF-sequence galaxies along the SSFR distribution.}
\label{histo}
\end{figure*}

\section{Sample selection and measurements}
\label{sample}

In order to obtain a meaningful census of star-forming galaxies on and off the MS,
one has to cover the SFR$-M_*$ plane down to low SFR and $M_*$
levels, and do so over a large area to include rare objects with the
highest SFRs.  We reach this goal by combining far-IR-selected (i.e.,
SFR-selected) and near-IR-selected (i.e., $M_*$-selected) star-forming samples in
the COSMOS and in GOODS-South fields, having both UV- and far IR-based
SFR determinations.  We first describe the datasets used, sample
selections and SFR and $M_*$ measurements.  We consider only galaxies
within the redshift range of $1.5<z<2.5$, based either on
spectroscopic or photometric redshifts. The datasets include in total
576 and 122 PACS-detected galaxies in the COSMOS and GOODS-South
fields, respectively, and 18,981 and 586 BzK-selected (cf. Daddi et
al. 2004) sources in the COSMOS and GOODS-South fields.

\subsection{SFR-selected Herschel samples}

Herschel-PACS observations cover 2.04 square degrees over the COSMOS
field down to a $5\sigma$ limit of 8 and 17~mJy at 100 and 160 $\mu$m,
respectively, well above confusion limits (Lutz et al. 2011). Blending
is not a major issue in the PACS data, and photometry was carried out
by PSF-fitting at 24$\mu$m prior positions.  The detection limits
correspond to $\sim 200\,M_\odot$~yr$^{-1}$ at $z\sim2$ (rising to 300$M_\odot~yr^{-1}$ at z=2.5).
We cross-matched over a common area of 1.73 square degrees the PACS
detections with the IRAC-selected catalog of Ilbert et al. (2010), so
to obtain UV-to-8$\mu$m photometry, accurate photometric redshifts and
stellar masses by SED fits as described in Rodighiero et
al. (2010b). At $z\sim2$ the mass completeness is granted above $\sim
10^{10}M_{\odot}$ (Ilbert et al. 2010).  IR luminosities (hence SFRs)
are derived from PACS fluxes using a set of empirical templates of
local objects (Polletta et al. 2007; Gruppioni et al. 2010) as
described in Rodighiero et al. (2010a).
If, instead, we adopt the templates from Chary \& Elbaz (2001), consistent
SFR estimates are obtained with no bias and a scatter of $\sim 0.15$~dex  
(that represents the typical error associated to our SFRs) .  
The PACS data in GOODS are deeper than those in COSMOS, reaching $5\sigma$ detection
levels of 1.8, 2.0 and 4.0~mJy at 70, 100 and 160 $\mu$m,
respectively (Berta et al. 2011).  The typical SFR limit at $z\sim2$ is
$\sim50M_\odot$~yr$^{-1}$.  For PACS data reduction, photometry,
far-IR based SFRs, and $M_*$ estimates we proceeded as for
the COSMOS field.

\subsection{Mass-selected BzK samples}

For the GOODS-S field we used the $K$-band selected sample of
star-forming BzK galaxies from Daddi et al. (2007), which reaches
$\sim M^* \sim10^{10}M_\odot$ over a 120 arcmin$^2$ field, with SFR
estimated from the UV rest-frame luminosity corrected for dust
reddening and taking advantage of a substantial fraction of
spectroscopic redshifts.  The deep BzK GOODS sample was complemented
with the larger statistics (in particular for the more massive
sources) provided by the BzK sample over the COSMOS field (McCracken et
al. 2010), where 
here masses and SFRs have been computed using the same procedure as in
Daddi et al. (2007), with typical errors of a factor $\sim2$. 

Use of UV-based SFRs allows one to reach
down to a few M$_\odot$~yr$^{-1}$ at $z\sim2$. Daddi et al. (2007)
performed detailed comparisons of UV-based SFRs with 70$\mu$m,
850$\mu$m, radio and X-ray based SFRs from stacking, finding good
agreement.  Pannella et al. (2009) used more than 30,000 BzK-selected
galaxies in COSMOS, finding that UV-based SFRs were slightly
underestimated compared to radio.
More recently Nordon et al. (2010) suggest that UV-based SFRs might actually be overestimated up to a
factor of 1.5-2 at $z\sim2$, using Herschel observations
(although see Wuyts et al. 2011a for a discussion of how this conclusion
depends on sample selections).  This relatively low level of uncertainty does not affect the main
conclusion of this paper.

\section{Results}
\label{results}

Fig.~\ref{selection} shows the SFRs versus stellar masses for the four
$1.5<z<2.5$ galaxy samples in this study. At first sight the
PACS-based SFR$-M_*$ relation for the COSMOS sample (red symbols)
bears little resemblance to the corresponding UV-based relation (black
points). PACS-based SFRs run almost flat with mass, as opposed to the
UV-based SFRs that increase almost linearly with mass. We maintain
that this different behavior is due to the different selection of the
two samples, one being SFR limited, the other being mass limited.  It
is immediately apparent from Fig. 1 that the vast majority of
BzK-selected galaxies of a few to several $10^{10}\,M_\odot$ are not
detected by PACS, implying that their SFR is lower than $200-300\,
M_\odot$yr$^{-1}$. Hence the intrinsic SFR$-M_*$ relation must be
steeper than suggested by the PACS data alone. That this must be the
case is indicated by the deeper PACS data over the GOODS field (cyan
points), which start populating to lower stellar masses the MS defined
by the BzK-selected sample with UV-based SFRs.
Still, the PACS SFR-selection cut is visible  at $\sim 60\,M_{\odot}/yr$,
and most BzK galaxies remain undetected below this limit.

The PACS-GOODS sample also allows to populate the region with excess SFRs
above the MS, something harder to do with UV-based SFRs due to
obscuration.  The combination of such datasets is clearly ideal to
obtain a statistical census of high-SFR galaxies as well as high SSFR
(see the insert of Fig.~\ref{selection}).
Indeed, half of PACS-detected sources over the
COSMOS field, either do not have a BzK counterpart, or their SFR is a
factor $\sim 4$ or more higher than their UV-based SFR. We interpret
this as evidence that most of the SF activity in these galaxies is
heavily dust-obscured in the UV.

In order to investigate the frequency and relative role  of
galaxies on and off the MS, we have derived the number density
distribution of galaxies in four  stellar mass bins as a function of their 
SSFR (see Fig.~\ref{histo}). We assume no redshift incompleteness
for the BzK galaxies over the $1.5<z<2.5$ range (see, e.g., McCracken et
al. 2010), whereas for PACS galaxies the flux
limits  imply different SFR limits as a function of redshifts. Therefore, 
 $1/V_{\rm max}$ corrections were computed with the same templates
used to derive the IR luminosities. This procedure implicitly assumes
no strong evolution of the number density of the population in the
probed redshift range. This is confirmed by  the average $V/V_{\rm max}$ 
being $0.504\pm0.023$ and
$0.56\pm0.05$,  for the COSMOS- and GOODS-PACS samples, respectively.
The SSFR distributions from the four samples agree within the errors
(Poisson) in the regions of overlap, providing an important
cross-check of the solidity of our approach.

In absence of a clear bimodality in Fig. \ref{selection},
the distribution functions shown in Fig.~\ref{histo} allow us to objectively define
MS outliers, hence starburst galaxies.
To this purpose, Gaussians functions are fitted to the BzK
distributions in the four mass bins, resulting in a nearly
constant $\sigma=0.24$~dex, slightly lower than reported in Daddi et
al. (2007).  Deviations from the Gaussian distributions start to be
clearly detected at SSFRs 0.6~dex above the average
(Fig.~\ref{histo}; or some 2.5$\sigma$), and we adopt this threshold to define on-sequence
and off-sequence galaxies.
Such deviations are less obvious in the highest mass bin.
We also note  that the peak positions of the Gaussians shift as a function of the bin central mass, with a slope of
$0.79\pm0.04$ in the log(SFR)--log($M_*$) relation{\footnote {The MS
best-fit is log(SFR)=-6.42+log(mass)*0.79}} (slightly shallower than
the 0.9 slope reported by Daddi et al. 2007).  It is worth emphasizing
the good agreement of the four independent samples in the common SSFR
bins, in particular for $M_*>10^{11}M_{\odot}$, where PACS-GOODS fully
samples the MS distribution around its peak.  In this case the UV and
IR SFR tracers define exactly the same Gaussian distribution.

\begin{figure}
\includegraphics[width=9cm]{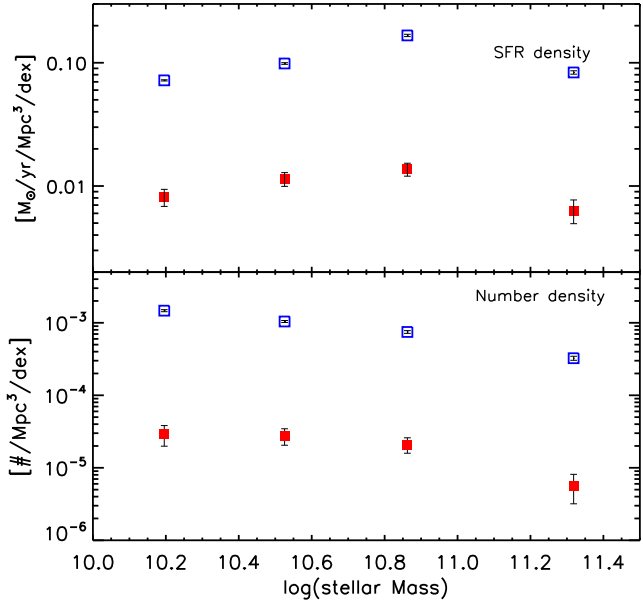}
\caption{
SFR density (top panel) and number density (bottom panel) of ON sequence sources (blue open squares) and OFF sequence sources (red filled squares),
in the four mass bins considered in Fig. \ref{histo}. Error bars are Poisson.
}
\label{density}
\end{figure}

\begin{figure}
\includegraphics[width=9cm]{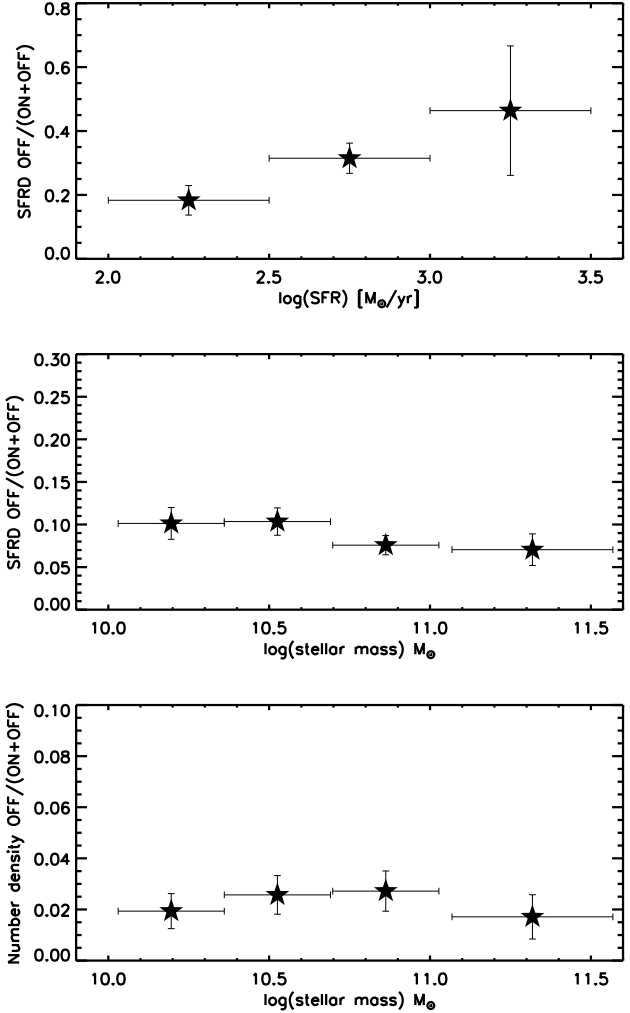}
\caption{
Contribution of OFF sequence galaxies to the total SFR density in different SFR bins (top panel) and
stellar mass bins (middle panel). In the bottom panel we also report the number density percentage of OFF sequence
sources.  Error bars are Poisson. 
}
\label{dist}
\end{figure}

From the histograms in Fig.~\ref{histo} we have estimated the relative 
contribution of on-sequence and off-sequence galaxies to the total comoving number
density and SFR density, in absolute (Fig.~\ref{density}) as well as relative
(Fig.~\ref{dist}) terms.  
The space density of galaxies in the highest mass bin is getting lower, as we are entering the exponentially
decaying part of the mass function.
The MS population dominates the SFR density at all masses, whereas  off-sequence starburst galaxies contribute
almost constantly $\sim 10\%$ of the total, or even less since some outliers may
actually be MS objects with exceptionally large errors (in either SFR or M$_*$).
If a top-heavy IMF were really to apply to starbursts, then this fraction would even be largely reduced.
The number density of off-sequence sources is also very small, varying
between 2 and 3$\%$ as a function of the stellar mass. Only in
SFR-limited samples (Fig.~\ref{dist}), off-sequence galaxies become
important representing $46\pm20$\% of the galaxies with
SFR$>1000$~$M_\odot$~yr$^{-1}$ and 20$\pm4$\% of those with
SFR$>100~$M$_\odot$~yr$^{-1}$. This suggests that even among SMGs (or luminous
Herschel selected populations) only a fraction of the galaxies are
strong MS outliers.

\section{Combining UV and IR SFR tracers}
An important aspect of this work concerns the combination of UV-based
and IR-based SFR indicators for the sources for which both data are available.
Indeed, about 70\%(80\%) of the COSMOS(GOODS) PACS sources have a BzK
counterpart.  For $\sim 30\%$ of these PACS-detected galaxies the
IR-based SFRs exceed the UV-based SFRs by factors of 4--10, that we
qualify as optically thick starbursts. For these sources the SFR(UV)
is totally unreliable.  However, these objects are almost all outliers
from the MS, and they represent only $\sim2$\% of the total BzK
population.  For the other objects, a direct comparison of UV-based
and far IR-based SFRs for GOODS BzK-selected galaxies shows reasonably
good agreement, consistently with the results of Nordon et al. (2010)
and Reddy et al. (2011 in prep.).

To further test the reliability of the SFR(UV), we analyzed the BzK
population with SFR(UV) in the same SFR range accessible by PACS in
the deep GOODS field.  We restricted our analysis to
$60<$SFR(UV)$<300M_{\sun}/yr$ objects in two mass bins: a)
$10.0<$log(M)$<10.75M_{\sun}$ and b) $10.75<$log(M)$<11.0M_{\sun}$, and
then stacked on the 160$\mu$m map the BzK PACS-undetected sources
combining them with PACS detections to get a mean far-IR SFR for the
BzK samples (as in Rodighiero et al. 2010a).  For the two mass bins the
result is: a) $<$SFR(UV)$>= 79\pm1M_{\sun}/yr$ and $<$SFR(PACS)$>=69\pm10M_{\sun}/yr$;
b) $<$SFR(UV)$>= 126\pm1M_{\sun}/yr$ and $<$SFR(PACS)$>=113\pm6M_{\sun}/yr$.  Reassuringly, in
the common mass and SFR ranges, the PACS and UV SFR are in
excellent agreement (with only $\sim$10\% of UV
overestimation in the highest mass bin that would not affect the
conclusion of this work).

Summarizing, we found that PACS and UV based SFRs are in good
agreement along the MS, while the off-sequence discrepancies (i.e., objects
for which UV can dangerously underestimate the actual SFR) are due to
dust obscuration at high SFR levels. These objects represent $\sim
2\%$ of star-forming galaxies at $z\sim 2$ and $M_*>10^{10}\,M_\odot$.

\section{The most extreme Star Forming sources and the obscured AGN activity}

We have shown that the distribution of SSFRs in $z\sim2$ massive
galaxies are largely Gaussian with $\sigma$ of 0.24~dex, and relatively
minor deviations only detected above $0.6~$dex of the main trend.
Although some of these high SSFR galaxies will just be normal MS
galaxies being scattered there by measurement errors, we
conservatively identify all of them as (merging-driven) starbursts. Of
course, also galaxies that are still on the MS can be
witnessing merging events, given that not all mergers are expected to
produce strong SFR enhancements (di Matteo et al. 2008), hence our
off-sequence galaxies can be seen as objects where a major merger
event has substantially boosted the SSFR, increasing it by a factor
of $\geq4$.

\begin{figure*}
\centering
\includegraphics[width=15cm]{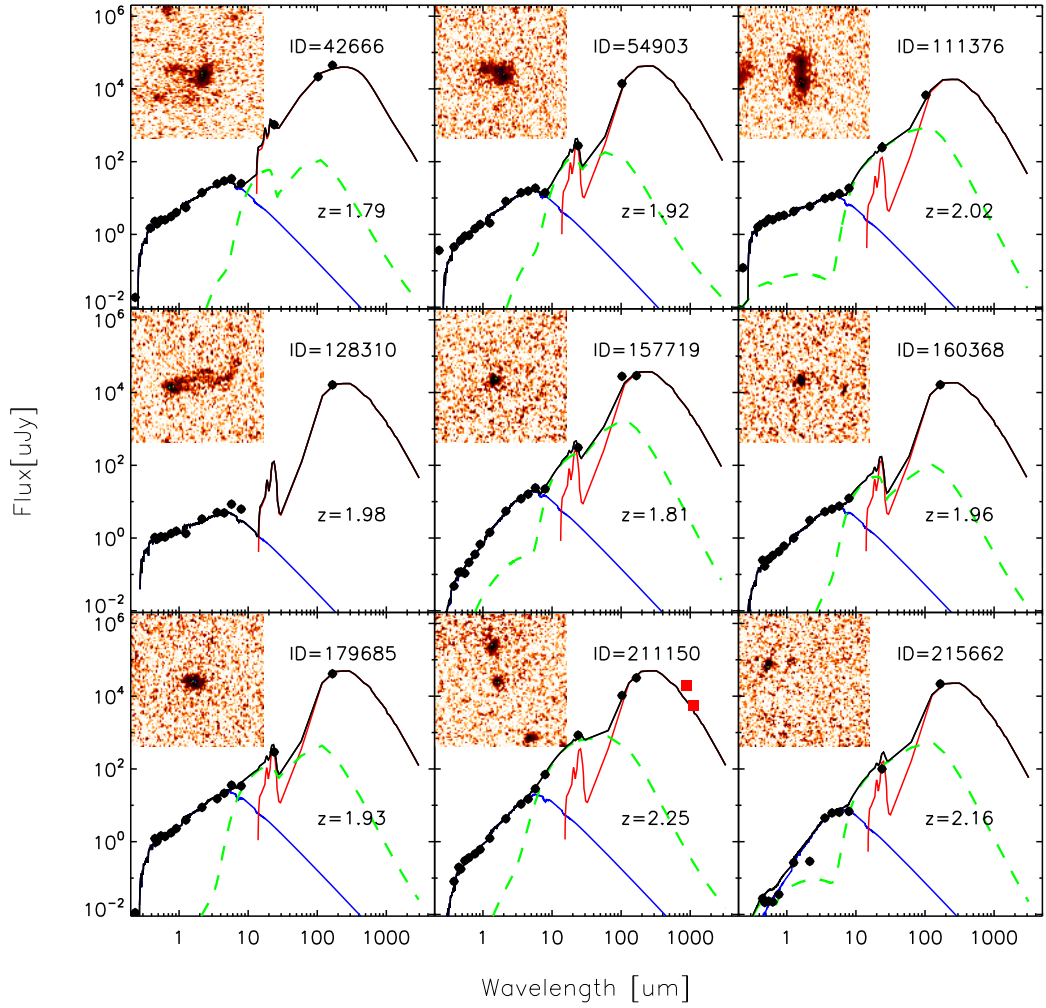}
\caption{Spectrophotometric properties of 9 out of 28 representative sources lying a factor 10 above the MS.
We performed a simultaneous three components SED fitting analysis
including stellar (blue), dusty torus (green) and starburst (red) emissions. The sum of the three components is also reported (black). 
A cutout of the ACS $i$ band ($5"\times5"$) is reported.
}
\label{SED}
\end{figure*}

In order to investigate the nature of off-MS galaxies we have focused
on the most extreme 28 objects, lying a factor 10 above the MS.  We
performed a simultaneous three components SED fitting analysis
including the emission of stars and star-heated dust,
and allowing also for the presence of a dusty torus (AGN), using the
method described in Fritz et al. (2006).
In Fig.~\ref{SED} we report an example of the resulting SED fitting obtained for 9 
representative objects, selected among the 28 sources.
The presence of an AGN is required in most cases (25 out of 28) to
reproduce the 24$\mu$m emission, while its contribution to the 
$L[8-1000\mu m]$ is always lower than 5-10\%, thus negligible.
None of the 16 galaxies observed by Chandra is
detected (Elvis et al. 2009), suggesting that the AGNs are extremely
obscured. These objects are probably similar to the Fe~$K_\alpha$
emitting Compton thick QSO at $z=2.5$ from Feruglio et al.  (2011),
that qualifies as a starburst galaxy according to our criterion.
Only 2 of our 28 sources fall in the AzTEC area (Scott et al. 2008),
and one of them is detected in the millimeter catalog. Fig.~\ref{SED}
also reports ACS-F775W cutouts for the example objects. A few objects are
consistent with advanced stage mergers (double components on the
$I$-band ACS COSMOS image, Koekemoer, A.M. et al. 2007), others are
quite compact (consistent with the analysis of Elbaz et al. 2011),
whereas some remain undetected. 
A more extended morphological analysis is postponed to a future paper 
(but see also Wuyts et al., 2011b).

\section{Discussion}

The redshift range $1.5<z<2.5$ corresponds to a $\sim2$~Gyr cosmic time interval,
hence our star-forming galaxies have on average spent in it $\sim1$ Gyr. Therefore,
the fact that only $\sim2$\% of the massive galaxies are off the main
sequence implies that on average each galaxy spends $\sim 20$~Myr in
such a phase.  This is actually much shorter than both the gas
depletion timescale and the (outer rotation) dynamical time in
starburst galaxies (Daddi et al. 2010; Genzel et al. 2010). This is
also much shorter than the expected duration of the SFR-excess phase
in mergers based on numerical simulations (e.g. di Matteo et al. 2008, Martig et al. 2010), 
where the duration of the phase with SFR in
excess of $>4$ over the pre-merger SFR is of order of 200-300~Myr
(Bournaud et al. 2011).  Most likely this is because only a fraction
of massive star-forming galaxies undergoes major mergers during this
time interval, and/or because most mergers do not produce a
substantial increase of the SFR (consistent with the simulations of di
Matteo et al. 2008).

All in all, our results quite clearly show that the merger-enhanced
SFR phases are relatively unimportant for the formation of stars in
$z\sim2$ galaxies, and probably at all redshifts given that $z\sim2$
is known to be the 'prime time' for SMGs (Chapman et al. 2005), and that
this is similar to what is observed in the local Universe
(e.g. the percentual contribution of starburst is very small, Sanders \& Mirabel, 1996).
Still, going through this merging-driven starburst phase may be a critical
phase for the transformation of star-forming galaxies into passive
ellipticals. Hence, we maintain that off-sequence galaxies are likely
to be crucial objects for our understanding galaxy formation and
evolution.

\begin{acknowledgements}
GR acknowledges support from ASI (Herschel Science Contract  I/005/07/0).         
ED acknowledges funding support from ERC-StG grant UPGAL 240039 and ANR-08-JCJC-0008.
PACS has been developed by a consortium of institutes led by MPE (Germany) and including UVIE 
(Austria); KU Leuven, CSL, IMEC (Belgium); CEA, LAM (France); MPIA (Germany); INAF- 
IFSI/OAA/OAP/OAT, LENS, SISSA (Italy); IAC (Spain). This development has been supported by the 
funding agencies BMVIT (Austria), ESA-PRODEX (Belgium), CEA/CNES (France), DLR (Germany), 
ASI/INAF (Italy), and CICYT/MCYT (Spain).
We thank the anonymous referee for a constructive report.
\end{acknowledgements}


\begin{thebibliography}{}

\bibitem[Berta et al.(2011)]{2011arXiv1106.3070B} Berta, S., et al.\ 2011, arXiv:1106.3070 
\bibitem[Bournaud et al.(2011)]{2011ApJ...730....4B} Bournaud, F., et al.\ 2011, \apj, 730, 4 
\bibitem[Brinchmann et al.(2004)]{2004MNRAS.351.1151B} Brinchmann, J., 
Charlot, S., White, S.~D.~M., Tremonti, C., Kauffmann, G., Heckman, T., \& Brinkmann, J.\ 2004, \mnras, 351, 1151 

\bibitem[Chapman et al.(2005)]{2005ApJ...622..772C} Chapman, S.~C., Blain, A.~W., Smail, I., \& Ivison, R.~J.\ 2005, \apj, 622, 772 
\bibitem[Chary \& Elbaz(2001)]{2001ApJ...556..562C} Chary, R., \& Elbaz, D.\ 2001, \apj, 556, 562 

\bibitem[Daddi et al.(2004)]{2004ApJ...617..746D} Daddi, E., Cimatti, A., 
Renzini, A., Fontana, A., Mignoli, M., Pozzetti, L., Tozzi, P., 
\& Zamorani, G.\ 2004, \apj, 617, 746 

\bibitem[Daddi et al.(2005)]{2005ApJ...631L..13D} Daddi, E., et al.\ 2005, \apjl, 631, L13 
\bibitem[Daddi et al.(2007)]{2007ApJ...670..156D} Daddi, E., et al.\ 2007, \apj, 670, 156 
\bibitem[Daddi et al.(2008)]{2008ApJ...673L..21D} Daddi, E., Dannerbauer, H., Elbaz, D., Dickinson, M., Morrison, G., Stern, D., \& Ravindranath, S.\ 2008, \apjl, 673, L21 
\bibitem[Daddi et al.(2009)]{2009ApJ...694.1517D} Daddi, E., et al.\ 2009, \apj, 694, 1517 
\bibitem[Daddi et al.(2010)]{2010ApJ...713..686D} Daddi, E., et al.\ 2010, \apj, 713, 686 
\bibitem[Daddi et al.(2010)]{2010ApJ...714L.118D} Daddi, E., et al.\ 2010, \apjl, 714, L118 
\bibitem[Di Matteo et al.(2008)]{2008A&A...492...31D} Di Matteo, P., Bournaud, F., Martig, M., Combes, F., Melchior, A.-L., \& Semelin, B.\ 2008, \aap, 492, 31 

\bibitem[Elbaz et al.(2007)]{2007A&A...468...33E} Elbaz, D., et al.\ 2007, \aap, 468, 33 
\bibitem[Elbaz et al.(2011)]{2011arXiv1105.2537E} Elbaz, D., et al.\ 2011, arXiv:1105.2537 
\bibitem[Elvis et al.(2009)]{2009ApJS..184..158E} Elvis, M., et al.\ 2009, \apjs, 184, 158 

\bibitem[Feruglio et al.(2011)]{2011ApJ...729L...4F} Feruglio, C., Daddi, E., Fiore, F., Alexander, D.~M., Piconcelli, E., \& Malacaria, C.\ 2011, \apjl, 729, L4 
\bibitem[Fritz et al.(2006)]{2006MNRAS.366..767F} Fritz, J., Franceschini, A., \& Hatziminaoglou, E.\ 2006, \mnras, 366, 767 

\bibitem[Geach et al.(2011)]{2011ApJ...730L..19G} Geach, J.~E., Smail, I., Moran, S.~M., MacArthur, L.~A., Lagos, C.~d.~P., \& Edge, A.~C.\ 2011, \apjl, 730, L19 
\bibitem[Genzel et al.(2010)]{2010MNRAS.407.2091G} Genzel, R., et al.\ 2010, \mnras, 407, 2091 

\bibitem[Goldader et al.(2002)]{2002ApJ...568..651G} Goldader, J.~D., Meurer, G., Heckman, T.~M., Seibert, M., Sanders, D.~B., Calzetti, D., \& Steidel, C.~C.\ 2002, \apj, 568, 651 
\bibitem[Gonz{\'a}lez et al.(2010)]{2010ApJ...713..115G} Gonz{\'a}lez, V., Labb{\'e}, I., Bouwens, R.~J., Illingworth, G., Franx, M., Kriek, M., \& Brammer, G.~B.\ 2010, \apj, 713, 115 
\bibitem[Gruppioni et al.(2010)]{2010A&A...518L..27G} Gruppioni, C., et al.\ 2010, \aap, 518, L27 

\bibitem[Ilbert et al.(2010)]{2010ApJ...709..644I} Ilbert, O., et al.\ 2010, \apj, 709, 644
\bibitem[Karim et al.(2011)]{2011ApJ...730...61K} Karim, A., et al.\ 2011, \apj, 730, 61 
\bibitem[Koekemoer et al.(2007)]{2007ApJS..172..196K} Koekemoer, A.~M., et al.\ 2007, \apjs, 172, 196 
\bibitem[Lutz et al.(2011)]{2011arXiv1106.3285L} Lutz, D., et al.\ 2011, arXiv:1106.3285 

\bibitem[Martig \& Bournaud(2010)]{2010ApJ...714L.275M} Martig, M., \& Bournaud, F.\ 2010, \apjl, 714, L275 
\bibitem[McCracken et al.(2010)]{2010ApJ...708..202M} McCracken, H.~J., et al.\ 2010, \apj, 708, 202 
\bibitem[Mihos \& Hernquist(1996)]{1996ApJ...464..641M} Mihos, J.~C., \& Hernquist, L.\ 1996, \apj, 464, 641 

\bibitem[Noeske et al.(2007)]{2007ApJ...660L..43N} Noeske, K.~G., et al.\ 2007, \apjl, 660, L43 
\bibitem[Nordon et al.(2010)]{2010A&A...518L..24N} Nordon, R., et al.\ 2010, \aap, 518, L24 

\bibitem[Pannella et al.(2009)]{2009ApJ...698L.116P} Pannella, M., et al.\ 2009, \apjl, 698, L116 
\bibitem[Peng et al.(2010)]{2010ApJ...721..193P} Peng, Y., et al.\ 2010, \apj, 721, 193 

\bibitem[Peng et al.(2011)]{2011arXiv1106.2546P} Peng, Y., Lilly, S.~J., Renzini, A., \& Carollo, M.\ 2011, arXiv:1106.2546 
\bibitem[Pilbratt et al.(2010)]{2010A&A...518L...1P} Pilbratt, G.~L., et al.\ 2010, \aap, 518, L1 

\bibitem[Poglitsch et al.(2010)]{2010A&A...518L...2P} Poglitsch, A., et al.\ 2010, \aap, 518, L2 
\bibitem[Polletta et al.(2007)]{2007ApJ...663...81P} Polletta, M., et al.\ 2007, \apj, 663, 81 


\bibitem[Renzini(2009)]{2009MNRAS.398L..58R} Renzini, A.\ 2009, \mnras, 398, L58 

\bibitem[Rodighiero et al.(2010)]{2010A&A...518L..25R} Rodighiero, G., et al.\ 2010, \aap, 518, L25 
\bibitem[Rodighiero et al.(2010)]{2010A&A...515A...8R} Rodighiero, G., et al.\ 2010, \aap, 515, A8 

\bibitem[Salim et al.(2007)]{2007ApJS..173..267S} Salim, S., et al.\ 2007, \apjs, 173, 267 
\bibitem[Sanders et al.(1988)]{1988ApJ...325...74S} Sanders, D.~B., Soifer, B.~T., Elias, J.~H., Madore, B.~F., Matthews, K., Neugebauer, G., \& Scoville, N.~Z.\ 1988, \apj, 325, 74 
\bibitem[Sanders \& Mirabel(1996)]{1996ARA&A..34..749S} Sanders, D.~B., \& Mirabel, I.~F.\ 1996, \araa, 34, 749 
\bibitem[Scott et al.(2008)]{2008MNRAS.385.2225S} Scott, K.~S., et al.\ 2008, \mnras, 385, 2225 
\bibitem[Scoville et al.(2007)]{2007ApJS..172....1S} Scoville, N., et al.\ 2007, \apjs, 172, 1 

\bibitem[Tacconi et al.(2008)]{2008ApJ...680..246T} Tacconi, L.~J., et al.\ 2008, \apj, 680, 246 
\bibitem[Tacconi et al.(2010)]{2010Natur.463..781T} Tacconi, L.~J., et al.\ 2010, \nat, 463, 781 

\bibitem[Takagi et al.(2008)]{2008MNRAS.389..775T} Takagi, T., Ono, Y., Shimasaku, K., \& Hanami, H.\ 2008, \mnras, 389, 775 

\bibitem[Wuyts et al.(2011)]{2011arXiv1106.5502W} Wuyts, S., et al.\ 2011, ApJ in press, arXiv:1106.5502 
\bibitem[Wuyts et al.(2011)]{2011arXiv1107.0317W} Wuyts, S., et al.\ 2011, submitted to ApJ, arXiv:1107.0317


\end{thebibliography}
\end{document}